\documentclass[
	 reprint,
	superscriptaddress,
	preprintnumbers,
	nofootinbib,
	amsmath,amssymb,
	aps,
	pra,
	dvipdfmx,
	floatfix,
	longbibliography
	]{revtex4-1}

\usepackage{graphicx,color,hyperref}
\usepackage{dcolumn}
\usepackage{bm}
\usepackage{subfigure}
\usepackage{ulem}
\usepackage{url}

\begin{document}

  \preprint{KUNS-2907, OU-HET 1126}

  \title{
A bound on energy dependence of chaos}

  \author{Koji Hashimoto}%
  \email{koji@scphys.kyoto-u.ac.jp}
  \affiliation{%
  Department of Physics, Kyoto University, Kyoto 606-8502, Japan
  }
  \author{Keiju Murata}%
  \email{murata.keiju@nihon-u.ac.jp}
  \affiliation{%
  Department of Physics, Nihon University, Sakurajosui, Tokyo
156-8550, Japan
  }
  \author{Norihiro Tanahashi}%
  \email{tanahashi@phys.chuo-u.ac.jp}
  \affiliation{%
    Department of Physics, Chuo University, Kasuga, Bunkyo-ku, Tokyo 112-8551, Japan
  }
  \author{Ryota Watanabe}%
  \email{watanabe@het.phys.sci.osaka-u.ac.jp}
  \affiliation{%
  Department of Physics, Osaka University, Toyonaka, Osaka 560-0043, Japan  }

  \begin{abstract}
We conjecture a chaos energy bound, 
an upper bound on the energy dependence of the Lyapunov exponent for any classical/quantum Hamiltonian mechanics and field theories. The conjecture states that
the Lyapunov exponent $\lambda(E)$ 
grows no faster than linearly in the total energy $E$
in the high energy limit. 
In other words, the exponent $c$ in $\lambda(E) \propto E^c \, (E\to\infty)$
satisfies $c\leq 1$.  
This chaos energy bound 
stems from thermodynamic consistency of out-of-time-order correlators (OTOC's) and
applies to any classical/quantum system with finite $N$ / large $N$
($N$ is the number of degrees of freedom) under plausible physical conditions on the Hamiltonians.
To the best of our knowledge the chaos energy 
bound is satisfied by any classically chaotic Hamiltonian system known, 
and is consistent with
the cerebrated chaos bound by Maldacena, Shenker and Stanford, which is for 
quantum cases at large $N$.
We provide arguments supporting the conjecture for generic classically chaotic billiards and
multi-particle systems.
The existence of the chaos energy bound may put a fundamental constraint on physical systems and the universe.

    \vspace*{10mm}
  \end{abstract}

  \maketitle



\noindent
{\bf Conjecture.}
--- The statement of our conjecture is as follows. For any Hamiltonian system with its Hermitian
Hamiltonian made by finite polynomials
in coordinate/field variables,\footnote{%
One can further restrict the domain of the spatial coordinates to some bounded region.
We then assume that all components of the extrinsic curvature on the boundary surface are finite.
}
the classical/quantum Lyapunov exponent
$\lambda(E)$ measured at energy $E$ in the high energy limit
satisfies the following upper bound on its power in the energy dependence, 
\begin{align}
c\leq 1 \quad \mbox{for} \quad \lambda(E) \propto E^c  \quad (E\to\infty)  .
\label{orera}
\end{align}
More precisely,
for a given system there exist $C>0$ such that $|\lambda(E)| \leq C E$ for any sufficiently large $E$.
For quantum systems, the quantum Lyapunov exponent is measured \cite{Maldacena:2015waa}
by out-of-time order correlators (OTOC's)~\cite{Larkin}.
We call \eqref{orera} {\it chaos energy bound}.

This conjecture\footnote{The conjecture was first mentioned in a footnote in \cite{Hashimoto:2016wme}.} is motivated by 
the well-definedness of the 
canonical ensemble for chaotic systems.
Suppose a quantum (classical) Hamiltonian system has a chaos, as the OTOC 
$-\langle E|[q(t),p(0)]^2|E\rangle$ (Poisson bracket $\{q(t),p(0)\}_{\rm P}^2$) grows as $\exp [2 \lambda(E) t]$.
Then the thermal Lyapunov exponent $\lambda_{\rm th} (T)$ is defined by
\cite{Hashimoto:2017oit,Hashimoto:2016wme}
\begin{align}
\lambda_{\rm th}(T) \equiv 
\frac{1}{t}\!\log\!\left[ \!-\!\! \int \!\!dE \,\rho(E) e^{-\beta E} \langle E| [q(t),p(0)]^2|E\rangle\right]
\label{lth}
\end{align}
for large $t$ (smaller than the Ehrenfest time), 
where $\rho(E)$ is the density of states\footnote{%
For the case of string theory, in spite of $\rho(E)$ being exponentially
growing in $E$, the convergence argument still works as long as the
temperature is lower than the Hagedorn temperature.
} and 
$\beta \equiv 1/T$ where $T$ is the temperature.\footnote{%
The definition \eqref{lth} differs from that in \cite{Hashimoto:2017oit,Hashimoto:2016wme} by a factor of $1/2$ since \eqref{lth} is what was used in the quantum large $N$ bound~\eqref{MSS}. 
}
The convergence of this integral~\eqref{lth} requires the chaos energy bound \eqref{orera}, therefore, 
the bound allows one to treat the system at finite temperature.
This argument for~\eqref{lth} 
applies no matter whether the system is quantum or classical, at finite $N$ or large $N$, where $N$ 
is the number of degrees of freedom of the system.

In the large $N$ limit one can replace $E$ by the temperature $T$.\footnote{%
See Sec.~I of the supplemental material for the large $N$ saddle point approximation.} 
Hence the chaos energy bound \eqref{orera} in the large $N$ limit leads to the {\it chaos temperature bound}
\begin{align}
c \leq 1
\quad \mbox{for} \quad \lambda_{\rm th} (T) \propto T^c  \quad (T\to\infty).
\label{oreraT}
\end{align} 
Let us remind the readers of 
the celebrated chaos bound conjectured by Maldacena, Shenker and Stanford (MSS)~\cite{Maldacena:2015waa} for large $N$ quantum systems, 
\begin{align}
	\lambda_{\rm th}(T) \leq 2\pi T/\hbar\,,
\label{MSS}
\end{align}
whose saturation is a discrimination diagnosis 
for existence of a black hole description in gravity dual.
We find that the quantum large $N$ bound \eqref{MSS} shows \eqref{oreraT}, thus the bound
\eqref{orera} (which can be applied to more generic systems\footnote{Note that the classical limit $\hbar\to 0$ of 
\eqref{MSS} does not lead to any bound.}) is consistent with \eqref{MSS}.
Furthermore, since the saturation of \eqref{MSS} needs the saturation of \eqref{orera}, 
we can further conjecture that any holographic quantum system dual to a black hole should saturate the chaos energy bound \eqref{orera}.

\vspace{3mm}

\noindent
{\bf Examples of systems.}
--- Any classically chaotic systems studied in literature
satisfy the chaos energy bound~\eqref{orera}, as far as we have checked.
Here we list some for the readers' reference.

First we note that an ordered phase in the high energy limit is allowed in many chaotic models including 
double pendulum and sigma models \cite{Hashimoto:2016wme}, meaning $c<0$ satisfying the chaos energy bound \eqref{orera}.\footnote{%
The H\'enon-Heiles system and particle/string motion around black holes \cite{Suzuki:1996gm,Hashimoto:2016dfz,Hashimoto:2018fkb}
may not allow the high energy limit.
}

For many-body systems, we are concerned with their largest Lyapunov exponent.
For the Fermi-Pasta-Uram $\beta$-model, an analytic formula for the largest Lyapunov exponent \cite{Casetti:1996zz}
gives $c \sim 1/4$ and the bound \eqref{orera} is satisfied.
For a large number of coupled rotors, the formula gives
$c = -1/6$~\cite{Casetti:1996zz}, which satisfies the bound \eqref{orera}.

As a field theoretic example, a chaotic string in AdS soliton geometry \cite{Ishii:2016rlk} shows 
$\lambda(E)\sim \log E$, consistent with the bound \eqref{orera}.
Thermalized fluids show $c=1/2$ \cite{Murugan:2019cmg}, satisfying \eqref{orera}.
Homogeneous Yang-Mills mechanics~\cite{Matinyan:1981dj} gives $c=1/4$~\cite{Chirikov:1981cm,Gur-Ari:2015rcq},
determined by the scaling.
In Yang-Mills theories on a lattice \cite{Muller:1992iw,Gong:1992yu,Gong:1993xu} 
(see also \cite{Biro:1994sh,Muller:1996br,Kunihiro:2010tg}),
the largest Lyapunov exponent
$\lambda(E) = (1/6)g^2 E$ (where $g$ is the 
coupling constant)
saturates the bound \eqref{orera}.

\vspace{3mm}

\noindent
{\bf General particle motion and billiards.}
--- We show that billiards and their generalization satisfy the bound~\eqref{orera}.
Classical billiards with a standard kinetic Hamiltonian $H = p^2/2m$ allows a particle motion
with the velocity $\dot{x}\propto \sqrt{E}$, thus any Lyapunov exponent of billiards,
which is
proportional to the inverse duration of hitting the boundary wall, satisfies $\lambda(E) \propto \dot{x} \propto \sqrt{E}$.
This exponent of the billiard, 
\begin{align}
c=1/2 \, ,
\end{align} 
is subject to the bound \eqref{orera}.

For a generalized billiard with a kinetic Hamiltonian $H = p^\gamma$, 
Hamilton's equation is $\dot{x}=\gamma p^{\gamma-1}\propto E^{(\gamma-1)/\gamma}$, thus the
Lyapunov exponent has a power 
\begin{align}
c=1-1/\gamma \, ,
\end{align}
which is less than 1 for any positive and finite $\gamma$.
Therefore, the bound \eqref{orera} is always satisfied.\footnote{%
Note that any negative $\gamma$ does not allow chaos to be defined, because 
the motion stops asymptotically.}$^{,}$\footnote{%
An example $H\sim e^{p^2}$ can violate the bound but
does not satisfy our polynomial assumption for $H$.
In fact, in quantum mechanics, $H \sim e^p, e^{p^2}$ are non-local and thus physically deserted.
See Sec.~II of the supplemental material for details about our assumptions on the physicality of the Hamiltonians.
}

The argument above is expected to apply
to any sparse many-body system of $N$-particles with a finite-range interaction,
as any potential boils down to a contact scattering at $E \to \infty$.
Then, the typical velocity of the particle is $v \sim \sqrt{E/N}$ for $\gamma=2$.
The scattering rate is proportional to $v$ and, thus, $\lambda(E)\propto \sqrt{E}$.

We can also show that billiards with softened walls,
which are particles in generic potentials, may obey
the bound~\eqref{orera}.
Consider a 2-dimensional system approximated by 
$H = p_1^2/2 + p_2^2/2 + x_1^m x_2^n = \dot{x}_1^2/2 + \dot{x}_2^2/2 + x_1^m x_2^n$,
where the last term is a dominant term in a generic potential $V(x)$ at $E\to\infty$. 
Here $m$ and $n$ are positive because the orbits for defining the chaos need to be bounded.\footnote{%
A class of hard-wall billiards is described by a limit $m,n \to \infty$.}
Then the following scaling symmetry\footnote{%
This scaling is a generalization of what is described in \cite{Chirikov:1981cm,Savvidy:1982jk,Biro:1994bi}.}
$t \to \alpha t$,  $x_i \to \alpha^{2/(2-m-n)}x_i$ and $H \to \alpha^{2(m+n)/(2-m-n)} H$ 
leads to an equation
$\lambda(E) t=\lambda (\alpha^{2(m+n)/(2-m-n)} E) \cdot \alpha t$, which is solved to give the exponent as 
\begin{align}
c=1/2-1/(m+n) \, .
\end{align}
Thus the general 2-dimensional classical mechanics satisfies the chaos energy bound \eqref{orera}.\footnote{In Sec.~III of the supplemental material
we provide calculations of the exponent $c$ for more generic Hamiltonians.}

\vspace{3mm}

\noindent
{\bf Speculations.}
--- 
The chaos energy bound \eqref{orera} in quantum field theories (QFT's) is naturally understood as follows.
First, notice that, although there can be many coupling constants in the theory, the one with the smallest mass dimension will be dominant at high energy. Let $g$ be such a coupling,\footnote{%
Even if there are multiple coupling constants with the smallest mass dimension, our argument still holds.} 
and denote its mass dimension as $d_g$. In other words, at high energy, the only dimensionful parameters at hand are $E$ and $g$. Then, using some dimensionless constants $a,\,b$ and $c$, the Lyapunov exponent should be written as
$\lambda = b E^c g^a$ with $a>0$ since the chaos should vanish when the nonlinearity goes away at $g=0$.
Since the Lyapunov exponent has mass dimension 1, the dimensional analysis determines $c$ as
\begin{align}
c = 1- a d_g \,.
\end{align}
Assuming that the QFT is consistent at high energy, 
the perturbative renormalizability requires $d_g\geq 0$, which means the chaos energy bound \eqref{orera}.
The renormalizability makes sure that no new structure emerges at higher energy scales, which is in accord with
the spirit of our polynomial assumption made for Hamiltonians. 

No matter whether the system is quantum or classical, at finite $N$ or large $N$,
the chaos energy bound \eqref{orera} 
applies. This universality may put a novel constraint on physical theories
and even the chaos of the universe.
For example, remember that the sum of the positive Lyapunov exponents is
the Kolmogolov-Sinai entropy growth rate. Since naively a subsystem with the dominant entropy production may
dominate the whole system, the fundamental theory of the universe may need to saturate the 
chaos energy bound, which could be a selection principle 
of a QFT 
dictating the universe.
The bound is saturated by QFT's with $d_g=0$, which are classically conformal theories.
Interestingly, the standard model of elementary particles is almost classically conformal \cite{Bardeen:1995kv}. 
Thus, the classical conformality as a principle of particle phenomenology \cite{Meissner:2006zh,Iso:2009ss}
can be motivated also from the saturation of the chaos energy bound \eqref{orera}.

\vspace{3mm}

\noindent
{Acknowledgment.}
---
The authors are grateful to S.~Sasa for prominent discussions and 
would like to thank G.~Shiu, S.~Iso, J.~Magan and S.~S.~Ray for valuable comments.
The work of K.~H.\ was supported in part by JSPS KAKENHI Grant No.~JP17H06462.
The work of K.~M.\ was supported in part by JSPS KAKENHI Grant No.~20K03976, 18H01214 and 21H05186.
The work of N.~T.\ was supported in part by JSPS KAKENHI Grant No.~18K03623 and 21H05189.
We acknowledge the hospitality at APCTP where part of this work was done.


\end{document}


\title{
    A bound on energy dependence of chaos  -- supplemental material}

  \author{Koji Hashimoto}%
  \email{koji@scphys.kyoto-u.ac.jp}
  \affiliation{%
  Department of Physics, Kyoto University, Kyoto 606-8502, Japan
  }
  \author{Keiju Murata}%
  \email{murata.keiju@nihon-u.ac.jp}
  \affiliation{%
  Department of Physics, Nihon University, Sakurajosui, Tokyo
156-8550, Japan
  }
  \author{Norihiro Tanahashi}%
  \email{tanahashi@phys.chuo-u.ac.jp}
  \affiliation{%
    Department of Physics, Chuo University, Kasuga, Bunkyo-ku, Tokyo 112-8551, Japan
  }
  \author{Ryota Watanabe}%
  \email{watanabe@het.phys.sci.osaka-u.ac.jp}
  \affiliation{%
  Department of Physics, Osaka University, Toyonaka, Osaka 560-0043, Japan  }


 \begin{abstract}
 In this supplemental material, we provide arguments on the following three points: I. Relation between our conjectured chaos energy bound and the chaos bound provided by Maldacena, Shenker and Stanford \cite{Maldacena:2015waa}, II. General constraints
 which needs to be imposed on
the Hamiltonians for the conjecture, III. Derivation of the scaling symmetry for Lyapunov exponents, and IV.
Derivation of the conjecture for a simple class of polynomial Lagrangians by using the scaling symmetry. 
 \end{abstract}

\maketitle

\section{Relation to the chaos bound of Maldacena, Shenker and Stanford}

The Maldacena-Shekner-Stanford (MSS) bound \cite{Maldacena:2015waa}
on the quantum Lyapunov exponent $\lambda_{\rm th}$ at temperature $T$ for large $N$ quantum field theories is 
\begin{align}
\lambda_{\rm th} \leq \frac{2\pi T}{\hbar} \, , 
\label{MSSb}
\end{align}
whose saturation provides a quantum system which may allow a dual black hole description.
On the other hand our chaos energy bound is
\begin{align}
c \leq 1 \quad \mbox{for} \quad \lambda(E) \propto E^c
\label{orerae}
\end{align}
for the energy dependence of the Lyapunov exponent $\lambda(E)$ in the high energy limit. 
Here in this section we study the consistency
between \eqref{MSSb} and \eqref{orerae}.
What we show is that the saturation of the MSS bound \eqref{MSSb} needs the saturation of our chaos energy bound \eqref{orerae} in the limit of the large number of degrees of freedom.\footnote{%
Note that in the classical limit $\hbar \to 0$ 
the MSS bound \eqref{MSSb} does not give any bound, while
our chaos energy bound applies also to generic classical chaos.
}

Basically our chaos energy 
bound \eqref{orerae} is for a fixed energy, meaning that it is in the micro-canonical ensemble, 
while the MSS chaos bound \eqref{MSSb} is for a fixed value of the temperature, in the canonical ensemble.
The standard relation between them for expectation values of a generic operator ${\cal O}$ is
\begin{align}
\langle {\cal O}\rangle_T 
= \int dE \, \rho(E) \langle E | {\cal O}|E \rangle e^{-\beta E}/Z
\, ,
\label{TE}
\end{align}
where $\rho(E)$ is the density of state for energy $E$, $\beta\equiv 1/T$ is the inverse temperature,
and $Z$ is the partition function,
\begin{align}
Z \equiv \int dE \, \rho(E)  e^{-\beta E} \, .
\end{align}
The chaos energy bound \eqref{orerae} is motivated by the finiteness of this relation for the choice of 
an OTOC as ${\cal O}$.

To study the relation between the MSS bound \eqref{MSSb} and our chaos energy bound \eqref{orerae},
let us remind the readers of the fact that in the standard situation with a large number of degrees of freedom $N \to \infty$
in multi-particle dynamics, there exists a relation between the total energy $E$ and the
temperature $T$ as 
\begin{align}
E=\gamma T\ .
\label{EgT}
\end{align}
The constant $\gamma$ satisfies $\gamma\sim N \gg 1$.
One can derive the relation by using
the steepest-descent approximation for the integral \eqref{TE} for the density of states $\rho(E) \propto E^\gamma$, 
\begin{align}
\rho(E) e^{-\beta E} \propto
\exp\left[
-\frac{1}{2\gamma T^2}(E-\gamma T)^2
\right]
\, ,
\label{SD}
\end{align}
where we have ignored the higher-order terms that give subleading contributions in the large $N$ limit.
If we introduce $\epsilon \equiv E/\gamma T$, the above expression becomes 
\begin{align}
\rho(E) e^{-\beta E} \propto
\exp\left[
-\frac{\gamma}{2}(\epsilon-1)^2
\right]
\, ,
\label{SD2}
\end{align}
which is highly localized at $\epsilon=1$ with the width $\Delta \epsilon = 1/\sqrt{\gamma} \sim 1/\sqrt{N}\ll 1$.
Then,
\eqref{TE} leads to 
\begin{align}
\langle {\cal O}\rangle_T  \simeq  \langle E | {\cal O}|E \rangle|_{E=\gamma T}  \, .
\label{O_T}
\end{align}
A naive application of this formula
to $\langle E | {\cal O}|E \rangle = e^{2\lambda(E)t}$ leads to a relation between the
thermal quantum Lyapunov exponent $\lambda_{\rm th}$ and the micro-canonical Lyapunov exponent $\lambda(E)$
given by
\begin{align}
\lambda_{\rm th} = 2\lambda(E=\gamma T)
\, .
\label{lamlam}
\end{align}
Using this, the saturation of the MSS bound \eqref{MSSb} means in the micro-canonical ensemble
\begin{align}
\lambda(E) = \frac{\pi}{\hbar \gamma}E
\, .
\label{EEE}
\end{align}
Therefore we conclude that
the saturation of the MSS chaos bound \eqref{MSSb} shows the saturation of our chaos energy
bound \eqref{orerae}.
Note that for this conclusion, the large $N$ limit is necessary for the validity of the steepest-descent method.\footnote{%
We can regard the thermal average \eqref{TE} as the Laplace transform. According to what we have shown, 
we expect that 
the inverse Laplace transform of $Z(\beta)e^{2\pi t/\beta}$ in the large $N$ limit, where $Z(\beta)$ is the partition function, should be given by $\rho(E) e^{2\lambda(E)t}$ with \eqref{EEE}. The dependence $\rho(E)\propto E^{\gamma}$ determines $Z(\beta)\propto \beta^{-\gamma-1}$. Thus, the inverse Laplace transform we consider is essentially
$$
	\frac{1}{2\pi i}\int_{a-i\infty}^{a+i\infty}d\beta~\beta^{-\gamma-1}e^{2\pi t/\beta}e^{\beta E}\, ,
$$
where $a>0$ is an arbitrary real number. 

We can evaluate this in the large $N$ limit in two different ways. The first way is to employ the residue theorem. Let $\gamma$ be some positive integer. Since the integrand vanishes in $|\beta|\to\infty$ and ${\rm Re}\beta<\infty$, the integral is given by the residue of $\beta^{-\gamma-1}e^{2\pi t/\beta}e^{\beta E}$ at the origin $\beta=0$, which is
$$
	\sum_n \frac{E^{n+\gamma}}{(n+\gamma)!}\frac{(2\pi t)^n}{n!}\equiv \sum_n e^{g(n)}\, .
$$
Suppose that the sum is dominated by $n=n^*\gg1$. In the large $N$ limit, we find $n^*=2\pi E t/\gamma$ and $e^{g(n^*)}\propto E^\gamma e^{n^*}$. This result is precisely what we have expected. Note that $E/\gamma\ll1$ is necessary for $n^*\gg1$ to hold for sufficiently large $t$. (When $\gamma$ is not an integer, there appears a branch cut in the $\beta$ plane, but we expect that, even in this case, a similar result holds.) 

The other way is as follows; writing the inverse Laplace transform as
$$
	\frac{1}{2\pi i}\int_{a-i\infty}^{a+i\infty}d\beta~\exp\left[-(\gamma+1)
\log
\beta +\frac{2\pi t}{\beta}+\beta E\right]\, ,
$$
we can evaluate this by the steepest-descent method. In the large $N$ limit, the second term in the exponent is negligible compared to the other terms. Then, the saddle point is found at $\beta \sim \gamma/E$. Since $a>0$, we can deform the integration path so that it goes through the saddle point $\beta=\gamma/E$. Then, the inverse Laplace transform is proportional to $E^\gamma e^{2\pi E t/\gamma}$ to the leading order in $1/N$. Again, this is as we have expected.
}

In the estimation above, we ignored the $N$ dependence of the operator expectation value $\langle E | {\cal O}|E \rangle$ for simplicity. In the following, we derive the condition on it by looking more precisely at
the validity of the steepest-descent method. Using $\epsilon \equiv E/\gamma T$ and \eqref{SD2}, the thermal expectation value \eqref{TE} takes the following form in the large $N$ limit,
\begin{align}
\langle {\cal O}\rangle_T 
\simeq
	\frac{\displaystyle \int d\epsilon~f(\epsilon)e^{-\frac{\gamma}{2}(\epsilon-1)^2}}{\displaystyle \int d\epsilon~e^{-\frac{\gamma}{2}(\epsilon-1)^2}}\,,
\end{align}
where $f(\epsilon)=\langle E | {\cal O}|E \rangle|_{E=\gamma T \epsilon}$. Previously, we naively applied the steepest-descent method to this, and found~\eqref{O_T}, i.e., $f(1)$.
However, obviously, we should be more careful about this manipulation. Expanding $f(\epsilon)$ around $\epsilon=1$ and performing the Gaussian integrations, we will find
\begin{align}
\langle {\cal O}\rangle_T 
\simeq	f(1)+\frac{1}{2\gamma}f''(1) +\cdots
	\,.
\end{align}
For the validity of the steepest-descent method, we want the first term to dominate the others. Assuming that the higher derivative terms are negligible, the condition is
\begin{align}
	|f(1)|\gg \frac{|f''(1)|}{\gamma}\,.
	\label{SDcondition1}
\end{align}
Now, let us consider the particular case of our concern, $f(\epsilon)=e^{2\lambda(N,E)t}$,
where we write the $N$-dependence of the Lyapunov exponent explicitly. In general, $\lambda(N,E)$ is written as
\begin{align}
	\lambda(N,E)\propto N^aE^c \sim N^{a+c}T^c\epsilon^c
\label{Lyapunov_N_dependence}
\end{align}
in the high energy and the large $N$ limit, where $a$ and $c$ are some constants. Then, the condition \eqref{SDcondition1} is
\begin{align}
	N \gg {\cal O}(N^{2(a+c)}T^{2c}t^2) + {\cal O} (N^{a+c}T^ct) \,.
\label{SDcondition2}
\end{align}
For $a+c\leq0$, this condition is satisfied for sufficiently large $t$ in the large $N$ limit, and the steepest-descent method is valid. 
For $a+c>0$, since the second term in the RHS becomes negligible compared to the first term, 
the sufficient condition for \eqref{SDcondition2} to be satisfied for sufficiently large $t$ is
\begin{align}
	a+c<\frac{1}{2}\,.
\label{SDcondition3}
\end{align}

Let us check whether our thermal averaging \eqref{lamlam} to find the micro-canonical form of the the MSS bound is consistent with the steepest-descent method. 
In fact, in the large $N$ limit, \eqref{EEE} satisfies \eqref{SDcondition3} for a sufficiently large $t$ since $a+c=0$. 
Therefore we conclude the correctness of our derivation of \eqref{EEE} and the fact that
the saturation of the MSS bound leads to the saturation of the chaos energy bound, in the large $N$ limit.

As an illustrating example,
let us consider non-interacting $N$ particles moving in a billiard. As we have seen in the main text of this paper, the Lyapunov exponent of a billiard system is determined by the energy per particle, 
$\lambda(N,E) \sim (E/N)^{1/2}$. 
Thus, $a+c=0$ in this case, and the condition \eqref{SDcondition3} is satisfied.

Another interesting example is a gas of $N$ particles interacting with each other by contact scattering in a $d$-dimensional box 
with volume $V$. Let the number density $\rho=N/V$ be some fixed constant.\footnote{%
If we instead fix $V$ and take $N\to\infty$, then the system becomes too dense, and the kinetic theory of gases is inapplicable.}
Note that the $(d-1)$-dimensional cross section $\sigma$ of the particle is independent of $N$. According to the kinetic theory of gases, the scattering rate is given by $\rho \sigma v$, where $v\propto (E/N)^{1/2}$ is the average velocity of the particle.
Since the Lyapunov exponent should be proportional to the scattering rate, we find $\lambda(N,E) \sim \rho \sigma (E/N)^{1/2}$. Thus, $a+c=0$ again, and the condition \eqref{SDcondition3} is satisfied, independent of the dimensionality of the box.


\section{General constraints on Hamiltonians}

Our conjecture on the chaos energy bound \eqref{orerae} should be valid only for reasonably physical Hamiltonian systems.
In stating the conjecture, we have assumed that the Hamiltonian
is Hermitian and is a ``proper" operator. 
We define ``proper" operators as those which are given by a finite product of
fundamental field operators in the Heisenberg representation. 
Here fundamental operators mean the field operators
appearing in the definition of the Lagrangian of the system and their momentum 
conjugates.\footnote{%
For example, this assumption on the Hamiltonians says that, in quantum/classical mechanics, 
the Hamiltonian is made by a finite sum of monomials of $p(t)$ and $x(t)$ such as $g \cdot(x(t))^m (p(t))^n$.
}
In this section, we study why this constraint on Hamiltonians is necessary, for reasonably
physical systems.

Above, by the ``reasonably physical'' systems we mean that their Hamiltonians 
satisfy all of the following conditions:
\begin{itemize}
\item[$\langle 1 \rangle$] Hermitian Hamiltonian which does not depend on time explicitly.
\item[$\langle 2 \rangle$] Hamiltonian whose time evolution is local when quantized.
\item[$\langle 3 \rangle$] Hamiltonian with no infinitely small structure. 
\end{itemize}
The first condition $\langle 1 \rangle$ is for the energy to be conserved, to define the micro-canonical
state. Let us see the importance of 
the second and the third conditions. 
As one can see below, Hamiltonians defined by proper operators
concretely satisfy 
$\langle 2 \rangle$ and $\langle 3 \rangle$.

The locality condition $\langle 2 \rangle$ is manifest
when one considers the following non-local
example in quantum mechanics: 
\begin{align}
H = e^{i p d} + e^{-i p d}
\, 
\label{Hcos}
\end{align}
which is the standard lattice Hamiltonian in quantum field theories with the lattice spacing $d$. 
We can see that the time evolution of this system has the non-locality of the size $d$ as follows.
Let us consider the initially localized wave function: $\psi(t=0,x)=\delta(x)$. The wave function at $t=\delta t$ is given by
\begin{equation}
\begin{split}
  \psi(t=\delta t, x) &= (1+ i H \delta t)\psi(t=0, x)\\
&= \delta(x) + i \{\delta(x+d)+\delta(x-d)\} \delta t\ .
\end{split}
\end{equation}
This shows the non-locality of time evolution of the wave function.
In the actual study of lattice field theories, normally the continuum limit $\delta \to 0$ 
is taken carefully so that no non-local
propagation (``lattice artifacts'') remains afterwards. 
The large $p$ behavior of \eqref{Hcos} is not described by any polynomial in $p$.

For quantum systems
it is reasonable to require the locality, in other words, the polynomial nature of the Hamiltonian
when it is written by momentum variables. Otherwise the chaos energy bound may be violated. 
In fact, we find that a billiard with the Hamiltonian 
\begin{equation}
 H=e^{p \delta}\ ,
\label{epd}
\end{equation}
easily saturates the chaos energy bound \eqref{orerae}: 
\begin{align}
c=1 \, ,
\, 
\end{align}
since the Lyapunov exponent in the billiard system is given as 
\begin{align}
\lambda(E) \propto \dot{x} \equiv \frac{\partial H}{\partial p} = e^{p\delta} \delta  \propto E 
\, .
\end{align}
One can find a slightly generalized example of a non-local Hamiltonian
\begin{align}
H = e^{p^n}
\, 
\label{epdn}
\end{align}
with $n>0$, which leads to
\begin{align}
\lambda(E) \propto E (\log E)^{(n-1)/n}
\, .
\end{align}
For $n>1$, this non-local example violates the bound.\footnote{%
The violation is due to the logarithmic part,
and the power part still satisfies the chaos energy bound \eqref{orerae}.
So the violation of the bound is marginal.}
We observe that non-locality in the quantized time evolution may easily saturate/violate the chaos energy bound.

Let us turn to the condition $\langle 3 \rangle$. It is related to $\langle 2 \rangle$ since a canonical transformation
in classical mechanics exchanging $p(t)$ and $x(t)$ brings the Hamiltonian such as~\eqref{epdn} to the one with 
a potential
\begin{align}
V = e^{x^n}
\, .
\end{align}
This potential is not a polynomial. 
As we see in Sec.~IV, we may understand this potential as the limiting case of a polynomial
with the power going to infinity, $x^\infty$, which tends to saturate the chaos energy 
bound \eqref{orerae}.

A similar consideration for \eqref{Hcos} leads to a potential, for example,  
\begin{align}
V = \cos(x^2)
\, ,
\end{align}
which allows infinitely small structure as one goes to large $x$. In this sense, 
the condition $\langle 3 \rangle$ needs to be imposed once the condition $\langle 2 \rangle$ is imposed.

As a violent example, consider a potential in two-dimensional mechanics, 
\begin{align}
V = 
\left(\frac{1}{\sin^2(x^2)} +\frac{1}{\sin^2(y^2)} 
\right)
+ \log (x^2 + y^2) \, .
\end{align}
The first term of this potential allows infinitely dense spikes as one goes away from the origin. 
Due to the second term, the potential bottom grows larger as well, such that at higher energy the particle feels
more dense spiky potential. Thus we expect that 
the resultant Lyapunov exponent may violate the chaos energy bound \eqref{orerae}.
In this manner, if we allow infinitely small structure for the potential in the Hamiltonian,
one may construct examples violating the chaos energy bound \eqref{orerae}.

One is also allowed to set a domain of configuration for the system, as in the case of billiards.
Standard billiards are defined by the shape of the domain in which the particle(s) can move,
and thus specified by the domain boundary and the reflection condition there.
Concerning the condition $\langle 3 \rangle$, we need to impose a condition for
the domain boundary; in order for the boundary to have no infinitely small structure, we
impose a condition that all components of the extrinsic curvature of the boundary
surface need to be finite.\footnote{%
This condition may be relaxed to include boundaries that are non-differentiable at finite number of corners, such as
Sinai billiards.
}

Note that this third condition is physically understood as the renormalizability in
quantum field theories.\footnote{%
Note that this does not apply to quantum mechanics.
}
Therefore, the condition is intuitively dealt with the dimension argument of the coupling constants in
quantum field theories, as we discussed in the main text of this paper.

We have a comment on the notion of the properness of operators. 
In this paper we argued that to make sense of the canonical ensemble the integral \eqref{TE}
needs to be convergent, which provides the chaos energy bound. 
For this argument,
we need to 
restrict ourselves to treat only proper operator ${\cal O}$ in the transform \eqref{TE}.
The reason is as follows.
Let us consider an operator ${\cal O}$ which is out-of-time ordered.
One may argue that, if one uses
$\log {\cal O}$ instead of ${\cal O}$ in the integrand of \eqref{TE}, the finiteness of the
integration
does not provide any bound for the Lyapunov exponent. 
However, we note that 
the composite operator giving an OTOC of fundamental field operators is proper,
while the logarithm of it needs infinite product of fundamental operators for its definition,
and thus the latter is not 
proper.\footnote{%
In perturbation theory, improper operators need infinite number of renormalization procedures to make them finite, 
for example by using normal ordering to eliminate operator contact terms. 
A simple illustrating 
example is ${\cal O}=e^{H^2}$ where $H$ is a Hamiltonian; this ${\cal O}$ does not give a
convergent \eqref{TE}. However $e^{H^2}$ needs infinite renormalization and thus an improper operator.
The conceptual origin of our conjecture stems from the consideration that 
the OTOC's which measure the chaos are at least proper operators.
We would like to thank S.~Sasa for discussions.
}
In this sense, we do not consider $\log {\cal O}$ for the integration in \eqref{TE}.
The ``properness" of operators is important for our OTOC's and our Hamiltonians.

\section{Scaling symmetry and the energy dependence of Lyapunov exponent}

In this section, we show the energy dependence of the Lyapunov exponent for Hamiltonian systems with a scaling symmetry, based on a formalism to evaluate the Lyapunov exponent for given dynamical systems.

Let us introduce a dynamical system described by an $n$-dimensional dynamical variable 
$\bm{\xi}$
and an equation of motion 
\begin{equation}
\dot{\bm{\xi}}^t = \mathbf{F}\left(\bm{\xi}^t\right)\,,
\label{xeq}
\end{equation}
where $t$ is the time variable and
$\bm{\xi}^t\equiv\bm{\xi}(t)$ is the solution at the time $t$ for a given initial condition $\bm{\xi}=\bm{\xi}^0$ at $t=0$.
For this dynamical system, the Lyapunov spectrum including the largest Lyapunov exponent may be evaluated as follows~\cite{Sandri,Parker,Oseledets}.
We first introduce the so-called variational equation given by
\begin{equation}
\dot{\bm{\Phi}}_t = D_{\bm{\xi}} \mathbf{F}\left(\bm{\xi}^t\right)\cdot\bm{\Phi}_t\,,
\quad
\bm{\Phi}_0=\mathbf{I}\,,
\label{Phieq}
\end{equation}
where $\mathbf{I}$ is an identity matrix 
and $D_{\bm{\xi}} \mathbf{F}\left(\bm{\xi}^t\right)$ is given by
\begin{equation}
D_{\bm{\xi}} \mathbf{F}\left(\bm{\xi}^t\right) = \left.\frac{\partial \mathbf{F}(\bm{\xi})}{\partial \bm{\xi}}\right|_{\bm{\xi}=\bm{\xi}^t}\,.
\label{DF}
\end{equation}
The matrix $\bm{\Phi}_t$ describes response of the solution $\bm{\xi}$ at time $t$ against perturbation to the initial condition $\bm{\xi}=\bm{\xi}^0$, that is, 
$\delta\bm{\xi}(t)=\bm{\Phi}_t\cdot\delta\bm{\xi}^0$. This 
$\bm{\Phi}_t$ is obtained by integrating (\ref{xeq}) and (\ref{Phieq}) simultaneously.
Then, it is known that eigenvalues of the matrix $\bm{\Phi}_t^*\cdot\bm{\Phi}_t$, where $\bm{\Phi}_t^*$ is the Hermitian adjoint of $\bm{\Phi}_t$, behave as $e^{2t\lambda_1}, \ldots, e^{2t\lambda_n}$, where $\lambda_1\geq\ldots\geq\lambda_n$ are Lyapunov exponents and $\lambda\equiv \lambda_1$ is the largest exponent, which becomes non-zero when the system is chaotic.
Roughly speaking,
the above formalism implies that the Lyapunov spectrum may be evaluated by taking an average of the real part of the eigenvalues of $D_{\bm{\xi}} \mathbf{F}\left(\bm{\xi}^t\right)$ over an trajectory, and in this sense the real parts of the eigenvalues of $D_{\bm{\xi}} \mathbf{F}\left(\bm{\xi}^t\right)$ may be regarded as a local version of the Lyapunov exponents.

The above formalism to evaluate the Lyapunov exponent $\lambda$ can be applied to a dynamical system described by a Hamiltonian $H$. For simplicity we assume that the number of degrees of freedom is two, while generalization to arbitrary number of degrees of freedom would be straightforward.
The equations of motion for this Hamiltonian are given by \eqref{xeq} with
\begin{equation}
\begin{aligned}
\bm{\xi}&=\left(x(t),y(t),p_x(t),p_y(t)\right)
\equiv \left(
\mathbf{x}, \mathbf{p}
\right)
\,,
\\
\mathbf{F}(\bm{\xi})
&=\left(
\frac{\partial H}{\partial \mathbf{p}}, -\frac{\partial H}{\partial\mathbf{x}}
\right)\,,
\end{aligned}
\end{equation}
and then $D_{\bm{\xi}} \mathbf{F}\left(\bm{\xi}^t\right)$ defined by \eqref{DF} is given by
\begin{equation}
D_{\bm{\xi}} \mathbf{F}\left(\bm{\xi}^t\right)
=
\begin{pmatrix}
\frac{\partial^2 H}{\partial\mathbf{x}\partial\mathbf{p}}
&
-\frac{\partial^2 H}{\partial\mathbf{x}^2}
\\
\frac{\partial^2 H}{\partial\mathbf{p}^2}
&
-\frac{\partial^2 H}{\partial\mathbf{p}\partial\mathbf{x}}
\end{pmatrix}\,.
\label{DF_H}
\end{equation}
Suppose that this system has the following scaling symmetry:
\begin{equation}
t \to \alpha t \, , \quad
x \to \alpha^{s_x} x \, , \quad
y \to \alpha^{s_y} y \, , \quad 
H \to \alpha^{s_L} H\,.
\label{scaling_x}
\end{equation}
Then the conjugate momenta scale as 
\begin{equation}
p_x = \frac{\partial L}{\partial \dot x}\to \alpha^{s_L-s_x+1} p_x\,, 
\quad
p_y = \frac{\partial L}{\partial \dot y}\to \alpha^{s_L-s_y+1} p_y\,.
\label{scaling_p}
\end{equation}
Let us examine the scaling property of the eigenvalues $\hat\lambda$ of $D_{\bm{\xi}} \mathbf{F}\left(\bm{\xi}^t\right)$ defined by \eqref{DF_H}.
By applying the scaling transformation \eqref{scaling_x} and \eqref{scaling_p} to $D_{\bm{\xi}} \mathbf{F}\left(\bm{\xi}^t\right)$, we find that its components behave as
\begin{align}
\left(\frac{\partial^2 H}{\partial\mathbf{x}\partial\mathbf{p}}\right)
&\to
\begin{pmatrix}
\alpha^{-1}\frac{\partial^2 H}{\partial x \partial p_x}&
\alpha^{-s_x+s_y-1} \frac{\partial^2 H}{\partial x \partial p_y}\\
\alpha^{-s_y+s_x-1} \frac{\partial^2 H}{\partial y \partial p_x}&
\alpha^{-1}  \frac{\partial^2 H}{\partial y \partial p_y}
\end{pmatrix}\,,
\notag
\\
-\left(\frac{\partial^2 H}{\partial\mathbf{x}^2}\right)
&\to
-\alpha^{s_L}
\begin{pmatrix}
\alpha^{-2s_x}    \frac{\partial^2 H}{\partial x^2}&
\alpha^{-s_x-s_y} \frac{\partial^2 H}{\partial x \partial y}\\
\alpha^{-s_x-s_y} \frac{\partial^2 H}{\partial y \partial x}&
\alpha^{-2s_y}    \frac{\partial^2 H}{\partial y^2}
\end{pmatrix}\,,
\notag
\\
\left(\frac{\partial^2 H}{\partial\mathbf{p}^2}\right)
&\to
\alpha^{-s_L}
\begin{pmatrix}
\alpha^{2s_x}    \frac{\partial^2 H}{\partial p_x^2}&
\alpha^{s_x+s_y} \frac{\partial^2 H}{\partial p_x \partial p_y}\\
\alpha^{s_x+s_y} \frac{\partial^2 H}{\partial p_y \partial p_x}&
\alpha^{2s_y}    \frac{\partial^2 H}{\partial p_y^2}
\end{pmatrix}\,,
\notag
\\
-\left(\frac{\partial^2 H}{\partial\mathbf{p}\partial\mathbf{x}}\right)
&\to
-\begin{pmatrix}
\alpha^{-1}\frac{\partial^2 H}{\partial x \partial p_x}&
\alpha^{s_x-s_y+1} \frac{\partial^2 H}{\partial x \partial p_y}\\
\alpha^{s_y-s_x+1} \frac{\partial^2 H}{\partial y \partial p_x}&
\alpha^{-1}  \frac{\partial^2 H}{\partial y \partial p_y}
\end{pmatrix}\,.
\label{DF_scaling}
\end{align}
The eigenvalues $\hat\lambda$ of $D_{\bm{\xi}} \mathbf{F}\left(\bm{\xi}^t\right)$ can be found by solving the eigenvalue equation
\begin{equation}
\det\left(
D_{\bm{\xi}} \mathbf{F}\left(\bm{\xi}^t\right)- \hat\lambda \mathbf{I}
\right)=0 \, .
\label{DF_EVeq}
\end{equation}
Using \eqref{DF_scaling}, it can be shown that the eigenvalue equation~\eqref{DF_EVeq} is invariant under the scaling transformation provided that $\hat \lambda$ scales as
\begin{equation}
\hat\lambda \to \alpha^{-1} \hat\lambda \, .
\label{scaling_hatlambda}
\end{equation}

The Lyapunov spectrum $\lambda_i$ including the largest Lyapunov exponent $\lambda$ is given by an average of $\hat\lambda$ calculated based on \eqref{Phieq}, hence $\lambda$ shows the same scaling as \eqref{scaling_hatlambda}, that is, $\lambda \to \alpha^{-1}\lambda$.
Since the energy of the system scales as $E\to \alpha^{s_L}E$,
from the scaling property of $\lambda$ it is concluded that 
\begin{equation}
\lambda \propto E^{-1/s_L}\,.
\label{E-dependence}
\end{equation}
In the next subsection, we argue that the exponent of the energy dependence of the Lyapunov exponent, $-1/s_L$, should not be greater than the unity, at least for a typical dynamical system under some sensible assumptions.

Equation~\eqref{Phieq} implies that the Lyapunov exponent $\lambda$ is never greater than the local Lyapunov exponent $\mathrm{Re}\hat\lambda$, and they can be equal to each other only when the direction of the eigenvector of $\bm{\Phi}_t$ corresponding to the largest Lyapunov exponent does not change in time evolution.
Such a consideration leads to an estimate of an upper bound on the Lyapunov exponent given by
\begin{equation}
\lambda\leq 
\frac{\int d^n\xi \max\bigl(\mathrm{Re}\hat\lambda(\bm{\xi})\bigr)\delta\left(H(\bm{\xi})-E\right)}
{\int d^n\xi \,  \delta\left(H(\bm{\xi})-E\right)},
\label{phase-space-average}
\end{equation}
where $\max\bigl(\mathrm{Re}\hat\lambda(\bm{\xi})\bigr)$ is the largest value of the real part of the eigenvalues of the matrix $D_{\bm{\xi}} \mathbf{F}\left(\bm{\xi}\right)$. The integral is taken over a constant-energy hypersurface $H(\bm{\xi})=E$ in the phase space, hence the right-hand side is the phase-space average of $\max\bigl(\mathrm{Re}\hat\lambda(\bm{\xi})\bigr)$.
The estimate \eqref{phase-space-average} would be accurate if the phase-space average taken in \eqref{phase-space-average} coincides with a long-time average over a trajectory, which is taken in \eqref{Phieq}.\footnote{%
It would be interesting to relate this argument to the strict upper bound of Lyapunov exponent
in random matrix theories \cite{Sutter}.}
For a dynamical system with a scaling symmetry~(\ref{scaling_x}),
the right-hand side of (\ref{phase-space-average}) shows a scaling same as \eqref{scaling_hatlambda},
which implies that the upper bound on $\lambda$ scales as $E^{-1/s_L}$.

Using \eqref{phase-space-average}
or the maximum value of $\mathrm{Re}\hat\lambda$ on a constant-$E$ hypersurface in the phase space,
one may estimate the energy dependence of the upper bound on $\lambda$ even for dynamical systems without scaling symmetry. We will report elsewhere in the future on such an estimate for general dynamical systems.

\section{The classical bound and general polynomial Lagrangians}

For classically chaotic systems, 
we shall provide some evidence for the chaos energy bound \eqref{orerae}.
In particular, we show that a generic two-dimensional classical mechanics shows $c<1$,
for the energy dependence of the Lyapunov exponent
$\lambda(E) \propto E^c$.

First, as one of the simplest examples, 
let us review the scaling symmetry studied in \cite{Akutagawa:2020qbj}\footnote{See \cite{Chirikov:1981cm} for an earlier study.} 
for the Lagrangian
of the two-dimensional classical system with dynamical variables $x(t)$ and $y(t)$,
\begin{align}
L = \frac12 \dot{x}^2 + \frac12 \dot{y}^2 - g x^2 y^2 \, 
\label{H22}
\end{align}
where $g$ is a positive constant. The conserved energy is given by
\begin{equation}
 E=\frac12 \dot{x}^2 + \frac12 \dot{y}^2 + g x^2 y^2 \ .
\end{equation}
This model allows the following scaling symmetry,
\begin{align}
&t \to \alpha t \, , \quad
x \to \alpha^{-1} x \, , \quad
y \to \alpha^{-1} y \, , \nonumber\\
&L \to \alpha^{-4} L \, ,\quad E\to \alpha^{-4} E\ .
\end{align}
As the Lyapunov exponent should satisfy
\begin{align}
\lambda(E)t = \lambda(\alpha^{-4}E) \alpha t \, ,
\label{Lyapscale}
\end{align}
we find 
\begin{align}
\lambda(E) \propto E^{1/4} \, ,
\end{align}
which means $c=1/4$. It satisfies the bound \eqref{orerae}.

Let us generalize the Lagrangian to the following form,\footnote{%
In the Lagrangian, we take absolute values for all the variables $x,y,\dot{x},\dot{y}$ implicitly.
We also omit the coefficient of each term in the Lagrangian because it is not important for following discussion.
}
\begin{align}
L = x^s y^{\tilde{p}} \dot{x}^a + x^{\tilde{q}} y^t \dot{y}^b - x^{\tilde{m}} y^{\tilde{n}}
\, .
\label{Hmng}
\end{align}
This is a general Lagrangian allowing arbitrary monomials
as the kinetic function and the potential.
It should be regarded as just a dominant part of some total Lagrangian
in the high energy limit, which means we pick up terms which are most divergent in $x, y \to \infty$.
So the following analysis can be applied to Lagrangians whose high energy behavior is governed by
\eqref{Hmng}.

By a variable redefinition, this Lagrangian 
can be cast into a simpler form
\begin{align}
L = y^p \dot{x}^a + x^q \dot{y}^b - x^m y^n
\, .
\label{Hmn}
\end{align}
The conserved energy is 
\begin{equation}
 E=(a-1)y^p \dot{x}^a + (b-1)x^q \dot{y}^b + x^m y^n\ .
\label{Emn}
\end{equation}
Let us show the chaos energy bound $c<1$ for the classical chaos of this Lagrangian.
Before looking for any scaling symmetry, we study the parametric conditions for the constants appearing 
in the Lagrangian \eqref{Hmn}.
In fact, for the system \eqref{Hmn} 
to exhibit any chaotic behavior, we need to require the following two
conditions.
\begin{itemize}
\item
Consistency of motion.

First we need to require 
\begin{align}
a>1 \, , \quad b>1
\label{ab}
\end{align}
for the powers of $\dot{x}$ and $\dot{y}$ in \eqref{Hmn}. 
Otherwise the motion stops asymptotically in time, as we see in the following.
Let us assume $a\leq 1$ and consider a simplified Lagrangian\footnote{%
The coefficient of the first term should be negative for $a\leq 1$. Otherwise, the energy is not bounded from below.
}
\begin{align}
L = -\dot{x}^a - x^2
\, .
\end{align}
The conserved energy can be rewritten as
\begin{align}
E = (1-a) \dot{x}^a + x^2 \, .
\, 
\end{align}
Note that the exponent $a$ needs to be positive, otherwise the motion suffers from infinite acceleration
($\dot{x}\to \infty$ while keeping a finite energy).

Let us show that $a \leq 1$ is inconsistent to define any chaos.
Using the conserved energy $E$, the motion is given by an integration of the equation 
\begin{align}
\dot{x} \propto (E-x^2)^{1/a}
\, .
\end{align}
Obviously the motion climbs up the potential until reaching $x=\sqrt{E}$. 
We can estimate the time duration to reach the point $x=\sqrt{E}$ as 
\begin{align}
t \propto \int^{\sqrt{E}} dx(E-x^2)^{-1/a}
\, .
\end{align}
The right hand side diverges for $0\leq a \leq 1$, meaning that the motion will take infinite time to climb the potential $x^2$, which shows that the chaos cannot be defined in the system. Therefore we need to require
\begin{align}
a > 1\, .
\end{align}
Therefore we conclude that, to define any chaos, 
the powers $a$ and $b$ appearing in the kinetic term must be larger than 1.

\item 
Boundedness of the potential.

The Lagrangian \eqref{Hmn} includes the potential term $x^m y^n$. If $p=q=0$, obviously we need to require
$m>0$ and $n>0$ for any orbit of the motion to be bounded. 
This boundedness of the potential depends on the 
definition of the variables, and we need 
to canonically normalize the kinetic term in~\eqref{Hmn}.
For this purpose we introduce new variables\footnote{%
One may wonder if the variable redefinition \eqref{XY} produces new terms other than $\dot{X}^a$ and $\dot{Y}^b$.
That is true, however, those new terms are expected to be subdominant in the high energy limit.
As an example, let us look at
$$
L = y^4 \dot{x}^2 + \dot{y}^2 - V(x,y) \, . 
$$
The variable redefinition $X \equiv y^2 x,\, Y=y$ leads to 
$$
L=\dot{X}^2 -4 Y^{-1} X \dot{X}\dot{Y} + 4 Y^{-2} X^2 \dot{Y}^2+ \dot{Y}^2 -V
\, 
$$
and the new terms, the second and the third terms, have negative power in $Y$, meaning that it should be subdominant
for large $Y$. 
}
\begin{align}
X \equiv y^{p/a} x \, , \quad Y \equiv x^{q/b} y \, .
\, 
\end{align}
Then the potential is written as
\begin{align}
x^m y^n 
= X^A Y^B 
\, 
\label{XY}
\end{align}
with\footnote{Here, we suppose that $ab-pq$ does not vanish. When it does, $X$ and $Y$ are not independent, i.e., not good variables.}
\begin{align}
 A \equiv \frac{a(bm-nq)}{ab-pq}  \, , 
\quad  
 B \equiv \frac{b(an-mp)}{ab-pq} \, .
\end{align}
Therefore, for this potential to give bounded orbits, we need to require
\begin{align}
A >0 \, , 
\quad 
B > 0 \, .
\label{bounded}
\end{align}
These are the conditions for the boundedness of the potential.
\end{itemize}
We keep in mind that the system \eqref{Hmn} is with the conditions \eqref{ab} and \eqref{bounded},
and consider a scaling symmetry. The symmetry of the Lagrangian 
$L$ \eqref{Hmn} is found as
\begin{align}
&t \to \alpha t \, , \quad
x \to \alpha^{s_x} x \, , \quad
y \to \alpha^{s_y} y \, , \nonumber\\ 
&L \to \alpha^{s_L} L \, , \quad E \to \alpha^{s_L} E
\, 
\end{align}
where\footnote{We assume the denominator of the $s$'s does not vanish; otherwise the system \eqref{Hmn} does not allow any nontrivial scaling symmetry.}
\begin{align}
s_x&=\frac{(p-n)(a-b)-a(p-b)}{(ab-pq)-(an-pm)-(bm-nq)}
\, ,
\\
s_y&=\frac{(m-a)(a-b)+a(a-q)}{(ab-pq)-(an-pm)-(bm-nq)}
\, ,
\\
s_L&= \frac{mb(a-p)+na(b-q)}{(ab-pq)-(an-pm)-(bm-nq)}
\, .
\end{align}
From the scaling symmetry, we have
$\lambda(E)t=\lambda(\alpha^{s_L}E) \alpha t$ and
find $\lambda(E)\propto E^{-1/s_L}$. 
Then, the energy exponent of the Lyapunov exponent is evaluated as
\begin{align}
c & = \frac{(pq-ab)+(an-pm)+(bm-nq)}{b(an-pm)+a(bm-nq)}
\nonumber
\\
& = \frac{-1 + A/a + B/b}{A+B} \nonumber \\
& < \frac{-1 + A + B}{A+B} \nonumber \\
& < 1 \, .
\, 
\end{align}
At the first and the second inequalities, we used 
conditions \eqref{ab} and \eqref{bounded}, respectively.
Therefore, the Lyapunov exponent $\lambda(E)\propto E^c$ of 
the general Lagrangian system whose high energy part is governed by \eqref{Hmng}
is shown to be subject to the bound $c<1$, and thus the chaos energy bound \eqref{orerae}.

The saturation of the bound \eqref{orerae} is achieved in the limit 
\begin{align}
A, B \to +\infty  \, , 
\quad
a, b \to 1 \, .
\end{align}
A simple example is
\begin{align}
p=q=0 \, , \quad
m,n \to +\infty \, , \quad a,b \to 1 \, ,
\end{align}
which is a billiard\footnote{%
Various shape of billiards can be constructed by using a potential, for example, 
$$
V = g_1 (a_1x)^{2m} + g_2 (a_2y)^{2m} + g_3 (a_3 x^2+ a_4 y^2)^m
$$
with $m \to\infty$. This potential makes a billiard ball to move only 
in the region $|x|<1/a_1$, 
$|y|<1/a_2$, and $a_3 x^2+ a_4 y^2 < 1$. Thus a variety of billiard walls can be implemented
in parametric limits of polynomial potentials.
Note that this kind of potential allows a scaling symmetry in which $x$ and $y$ scale
in the same manner, thus the scaling argument described above can be applied to a large class of
billiards.  
} 
with a modified kinetic term $p^\gamma$ with $\gamma \to +\infty$.

\begin{acknowledgments}
We would like to thank S.~Aoki, S.~Heusler and S.~Sugimoto for valuable comments.
\end{acknowledgments}

\vfill
